\documentclass[11pt]{article}
\usepackage{geometry}                
\usepackage[sort,numbers]{natbib}
\usepackage{graphicx}
\usepackage{amssymb}
\usepackage{amsmath}
\makeatletter
\let\saved@includegraphics\includegraphics
\AtBeginDocument{\let\includegraphics\saved@includegraphics}
\renewenvironment*{figure}{\@float{figure}}{\end@float}
\makeatother

\title{A reproducible effect size is more useful than an irreproducible hypothesis test to analyze high throughput sequencing datasets}

\author{Andrew D. Fernandes\,$^{\text{1}}$ , Michael T.H.Q. Vu\,$^{\text{1}}$, Lisa-Monique Edward\,$^{\text{1}}$, \\
Jean M. Macklaim,$^{\text{2}}$ and Gregory B. Gloor\,$^{\text{1}*}$}

\begin{document}

\maketitle

\noindent{$^{\text{\sf 1}}$Department of Biochemistry, University of Western Ontario, London, N6A 5C1, Canada and \\
\noindent$^{\text{\sf 2}}$DNA GenoTek, Ottawa.}\\
\noindent{$^\ast$To whom correspondence should be addressed.}

\abstract{
\noindent\textbf{Motivation:} High throughput sequencing is analyzed using a combination of null hypothesis significance testing and ad-hoc cutoffs. This framework is strongly affected by sample size, and is known to be irreproducible in underpowered studies, yet no suitable non-parameteric alternative has been proposed. \\
\textbf{Results:} Here we present implementations of non-parametric standardized median effect size estimates,  $\mathcal{E}_{d}$, for high-throughput sequencing datasets. Case studies are shown for transcriptome and amplicon-sequencing datasets.  The  $\mathcal{E}_{d}$ statistic is shown to be more reproducible and robust than p-values and requires sample sizes as small as 5 to reproducibly identify  differentially abundant features.\\
\textbf{Availability:} Source code and binaries freely available at:\\
 https://bioconductor.org/packages/ALDEx2.html, omicplotR, and\\  https://github.com/ggloor/CoDaSeq.\\
\textbf{ggloor@uwo.ca}{ggloor@uwo.ca}\\
\textbf{Supplementary information:} Supplementary data and code will be available when published.

\section*{Introduction}

High throughput sequencing (HTS) datasets for transcriptomics, metagenomics and 16S rRNA gene sequencing are high dimensional, commonly conducted with pilot-scale  experiments and analyzed using a null hypothesis significance testing framework \citep{Schurch:2016aa}. Much effort has been spent identifying the best approaches and tools to determine what is `significantly different' between groups \citep{Soneson:2013,Schurch:2016aa}, but the answer seems to depend on the specific dataset and associated model parameters \citep{Thorsen:2016aa,hawinkel2017,Weiss:2017aa}. As commonly conducted the investigator determines what is `significantly different' using a null hypothesis significance approach and then decides what level of difference is `biologically meaningful' among the `significantly' different features. Graphically, this approach is  represented by the Volcano plot \citep{Cui:2003aa} where the magnitude of change (difference) is plotted vs the p-value.  One under-appreciated consequence of pilot scale research is that features with significant p-values   will have dramatically exaggerated  apparent effect sizes \citep{Halsey:2015aa}. This explains in part why so many observations  of apparent large effect fail to replicate in larger datasets \citep{Ioannidis:2005aa}. In fact, both p-values and absolute differences are poor predictors of which differences would reproduce if the experiment were conducted again \citep{Cumming:2008aa,Halsey:2015aa}



As many have pointed out, p-values are not useful proxies for biological relevance since p-values are designed, colloquially speaking, to estimate the likelihood of no true difference; p-values are not a test that the alternate hypothesis is true. They can only be used to estimate false-discovery rates if the p-value is calculated from a distribution that is appropriate for the experimental data, if reasonable estimates for the statistical power exist and a if reasonable estimate of the a priori probability that the null hypothesis is false \citep{Colquhoun:2014aa, Halsey:2015aa}. Simply put, p-values can only be used to test \emph{if} there is no difference between groups, not to measure the \emph{magnitude} of change between groups \citep{coe2002s,Colquhoun:2014aa}. The tension between the information that p-values provide and what the investigator needs is why magnitude of change cutoffs \citep{Cui:2003aa}, or other ad-hoc methods are used when deciding what is biologically relevant. Null-hypothesis significance based testing methods  also have the property that the number of significant features identified  is affected by the number of samples being compared. Thus leads to the concept of statistical power, where the experiment is designed such that statistical power is prioritized over biological significance. 

On the other hand, a standardized effect size addresses the issues of interest to the biologist:``what is reproducibly different?'' or ``would I identify the same true positive features as differential if the experiment were repeated?"  \citep{coe2002s,shinichi:2004,Colquhoun:2014aa,gloor:effect}. Standardized effect size statistics start from the assumption that there is a difference, but that the difference can be arbitrarily close to zero. Unfortunately,  standardized effect size metrics are not routinely used when analyzing HTS datasets, and one potential barrier to their use is that parametric effect size statistics may not be suitable for  HTS datasets.  

Here we introduce a simple non-parametric standardized effect size statistic for distributions, $\mathcal{E}_{d}$, that is implemented in the ALDEx2, omicplotR  and CoDaSeq R packages.  The $\mathcal{E}_{d}$ statistic has been used in both meta-transcriptome and microbiome studies, for example see \citep{macklaim:2013, bian:2017}, and has been shown to give remarkably reproducible results even with extremely small sample sizes \citep{nelson:2015vaginal}.   $\mathcal{E}_{d}$  has a near monotonic relationship with p-values (Supplement Figure 1). However, it is unknown how  $\mathcal{E}_{d}$  compares with  p-values or other effect size estimates, how many samples are required, and its sensitivity and specificity characteristics. 

\section*{Approach and Methods}

High throughput sequencing (HTS) machines output thousands to billions of `reads', short nucleotide sequences that are derived from a DNA or RNA molecule in the sequencing `library'. The library is a subset of the nucleic acid molecules that have been collected from an environment and made compatible with a particular HTS platform. The HTS instruments deliver these reads as integer `counts' per genomic feature---gene, location, etc. However, the counts are actually a single proxy for the probability of observing the particular read in a sample under a repeated sampling model; this is clear since technical replicates of the same library return different counts. The difference between technical replicates is consistent with multivariate Poisson sampling \citep{fernandes:2013, gloorAJS:2016} The probability estimate is delivered by the instrument as an integer representation of the probability multiplied by the number of reads  \citep{fernandes:2013, gloorAJS:2016}. Thus, the data returned by HTS are a type of count compositional data, where only the relationships between the features have meaning \citep{aitchison:1986, Lovell:2015,fernandes:2014,gloorFrontiers:2017,Kaul:2017aa}. 

The ALDEx2 tool uses a combination of probabilistic modelling and compositional data analysis to determine the features that are different between groups, where that difference is insensitive to random sampling. Technical replicate variance estimation and conversion of the count data to probabilities is accomplished by Monte-Carlo sampling from the Dirichlet distribution \citep{fernandes:2013, gloorAJS:2016}, which is conveniently also the conjugate prior for the multivariate Poisson process. The differences between features is linearized by applying a log-ratio transformation to the Dirichlet Monte-Carlo realizations and analyzed according to the rules of compositional data analysis \citep{aitchison:1986,fernandes:2013,Tsilimigras:2016aa,gloorFrontiers:2017}.
	
Starting with two vectors $\vec{a}$ and $\vec{b}$  that correspond to the concatenated log-ratio transformed Dirichlet Monte-Carlo realizations of a feature in two groups, we need a method to determine the standardized effect size  for the log-ratio transformed posterior probability estimates;  that is, the difference between groups relative to an estimate of within-group dispersion. Since these posterior distributions can have heavy tails, be multimodal, and be skewed, any useful statistic should be insensitive to even extreme non-Normality and provide sensible answers even if the posterior picture distributions are almost Cauchy in one or both groups \citep{fernandes:2013}. Below and in the Supplement we define the properties of the approach used. 

Cohen's d is a parametric standardized effect size for the difference between the means of two groups, and a general formulation is given in Equation~\ref{eq:cohen}. Cohen's d is the difference between the means of the two distributions divided by the pooled standard deviation, denoted as \(\sigma_{a,b}\). However, this metric depends upon the data being relatively Normal, which cannot be guaranteed for HTS data. 

\begin{equation}
\mathrm{d} = \frac{\mathrm{mean}(\vec{a} )- \mathrm{mean}(\vec{b})}{\sigma_{a,b}}
\label{eq:cohen}
\end{equation}

We can define a non-parametric  \emph{difference} vector  in Equation~(\ref{eq:diff}) as the signed difference between the two groups. We can further define a non-parametric  \emph{dispersion} vector as in Equation~(\ref{eq:disp}), where the notation $\boldsymbol{\rho}\vec{a}$ indicates one or more random permutations of the vector. Finally, we can define an \emph{effect} vector as in Equation~(\ref{eq:ff}) that is the ratio of these two non-parametric statistics. 

\begin{equation}
\vec{\delta} = \vec{a} - \vec{b}
\label{eq:diff}\vspace*{-10pt}
\end{equation}

\begin{equation}
\vec{\sigma} = max \{ \lvert \vec{a} - \boldsymbol{\rho} \vec{a}  \rvert ,\lvert \vec{b} -\boldsymbol{\rho} \vec{b} \rvert \}
\label{eq:disp}\vspace*{-10pt}
\end{equation}

\begin{equation}
\vec{\varepsilon} = \frac{\vec{\delta}}{\vec{\sigma}}
\label{eq:ff}\vspace*{0pt}
\end{equation}

Taking the median of $\vec{\delta}, \vec{\sigma}$ and $\vec{\varepsilon}$ returns a robust estimate of the central tendency of these statistics ($\tilde{D}$, MMAD (median of the maximum absolute deviation), and $\mathcal{E}_{d}$), and these are the `diff.btw', `diff.win' and `effect' statistics reported by ALDEx2. $\tilde{D}$ is the same as the difference between the means or the difference between medians in a Normal distribution as shown in Supplementary Figure  2. The MMAD metric is novel and the Supplement shows it has a Gaussian efficiency of 52\%, a breakdown point of 20\% (Supplementary Figure 3), and is 1.42 times the size of the standard deviation on a Normal distribution. The $\mathcal{E}_{d}$ statistic is a standardized effect size and is approximately 0.7 of Cohen's D when comparing the difference between two Normal distributions.  This is simply a Monte-Carlo estimate of a function of the respective random variables. Below and in Supplementary Figure 4 we show that this metric returns sensible values even with a Cauchy distribution.

We used  simple simulated datasets to determine baseline characteristics in a number of different distributions. Then we use the data  from a highly replicated RNA-seq experiment \citep{Schurch:2016aa} and examined 100 random subsets of the data with between 2 and 40 samples in each group. For each random subset we collected the set of features that were called as differentially abundant at thresholds of $\mathcal{E}_{d} \ge 1$, or with an expected Benjamini-Hochburg adjusted p-value of $\le 0.1$ calculated using either the parametric Welch's t-test, or the non-parametric Wilcoxon test in the ALDEx2 R package. These are output as `we.eBH' and 'wi.eBH' by the ALDEx2 tool. These were compared to a `truth' set determined by identifying those features that were identified in all of 100 independent tests of the full dataset with outliers removed using the same tests and cutoffs. Note that this is simply a measure of consistency and is congruent with the approach taken in \citep{Schurch:2016aa}.

The median values and 99\% confidence intervals for the number of true positives, false positives, true negatives and false negatives were tabulated and plotted in Fig. 1.

\section*{Results and Conclusions}

Measuring differential abundance in high throughput sequencing datasets is difficult for a variety of reasons. First, almost all experiments are underpowered. Second, the true distribution of the data is unknown. Third, when sample sizes are large almost all features are identified as `significantly different' by null hypothesis significance testing frameworks. 

We began by examining behaviour of   $\mathcal{E}_{d}$ and Cohen's~d statistics in simulated distributions as shown in Figure \ref{fig:01}. For this, we generated 100000 distributions for sample sizes  of between 2 and 40 in each group, and calculated the median and 99\textsuperscript{th} percentile effect size. When there is no difference between groups as illustrated in the top two panels, all four distributions had a median effect size of 0 at all sample sizes. However, the  99\textsuperscript{th} percentiles were different between the two statistics. We observe that the 99\textsuperscript{th} of the $\mathcal{E}_{d}$ metric increases more rapidly than does Cohen's~d at very small sample sizes, and that the  99\textsuperscript{th} percentile of the effect size of the Cauchy distribution is always somewhat less than the other three distributions for the $\mathcal{E}_{d}$ metric. In contrast, the  99\textsuperscript{th} percentile of the Cauchy distribution is much larger than the other distributions at low sample sizes, and tends to become smaller than the other distributions at large sample sizes. This behaviour is perhaps not surprising since the parametric Cohen's~d effect size should be undefined in a Cauchy distribution. 

\begin{figure}[tpb]
\centerline{\includegraphics[scale=0.50]{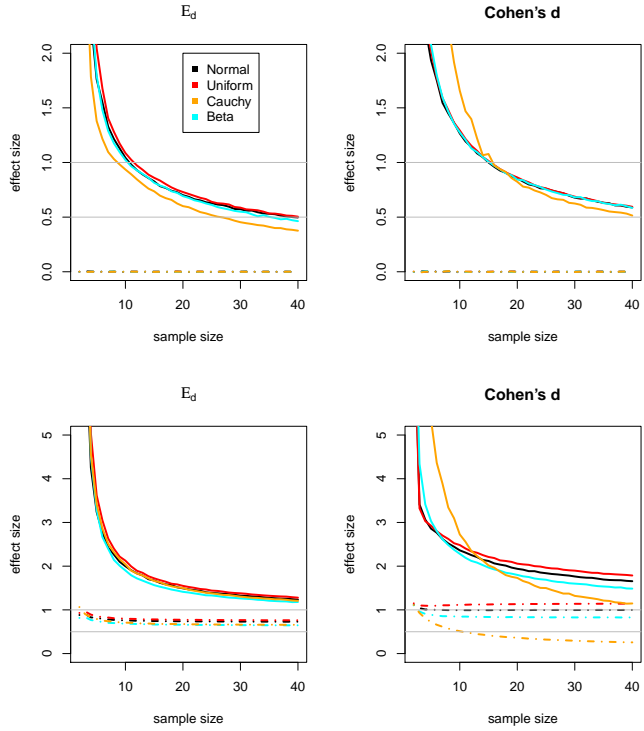}}
\caption{Characteristics of $\mathcal{E}_{d}$ and when detecting differential features in two groups in random Normal, random uniform, random Cauchy and a heavily skewed random $\beta$ distribution as a function of the per group sample size. The top panels show the behaviour of the $\mathcal{E}_{d}$ and Cohen's~d effect sizes when the two groups are drawn from the same distribution. The simulation included one-hundred thousand random distributions for each. The bottom two panels show the behaviour when there is a difference between groups. Single examples of the distributions are shown in the Supplement. The dashed lines show the median effect size, the solid lines show the 99\textsuperscript{th} percentile of the statistics.  }
\label{fig:01}
\end{figure}

The bottom two panels in Figure \ref{fig:01} show the behaviour when there is a small difference between groups; for reference; Supplementary Figure 5 shows single examples of the distributions used  for comparison.  Again, we can see that the distribution of the median and 99\textsuperscript{th} percentile difference at low sample sizes and are remarkably similar when using the  non-parametric $\mathcal{E}_{d}$ effect size metric, but diverge substantially when using the parametric Cohen's~d. Thus, we conclude that if the data were drawn from a Normal distribution that Cohen's~d would be preferred. However, given that distributions from actual sequencing datasets are unknown, and can be nearly Cauchy, multimodal or skewed \citep{fernandes:2013},  the non-parametric $\mathcal{E}_{d}$ effect size statistic gives a more stable and robust estimate of the standardized difference between distributions than does Cohen's~d.

\begin{figure}[tpb]
\centerline{\includegraphics[scale=0.5]{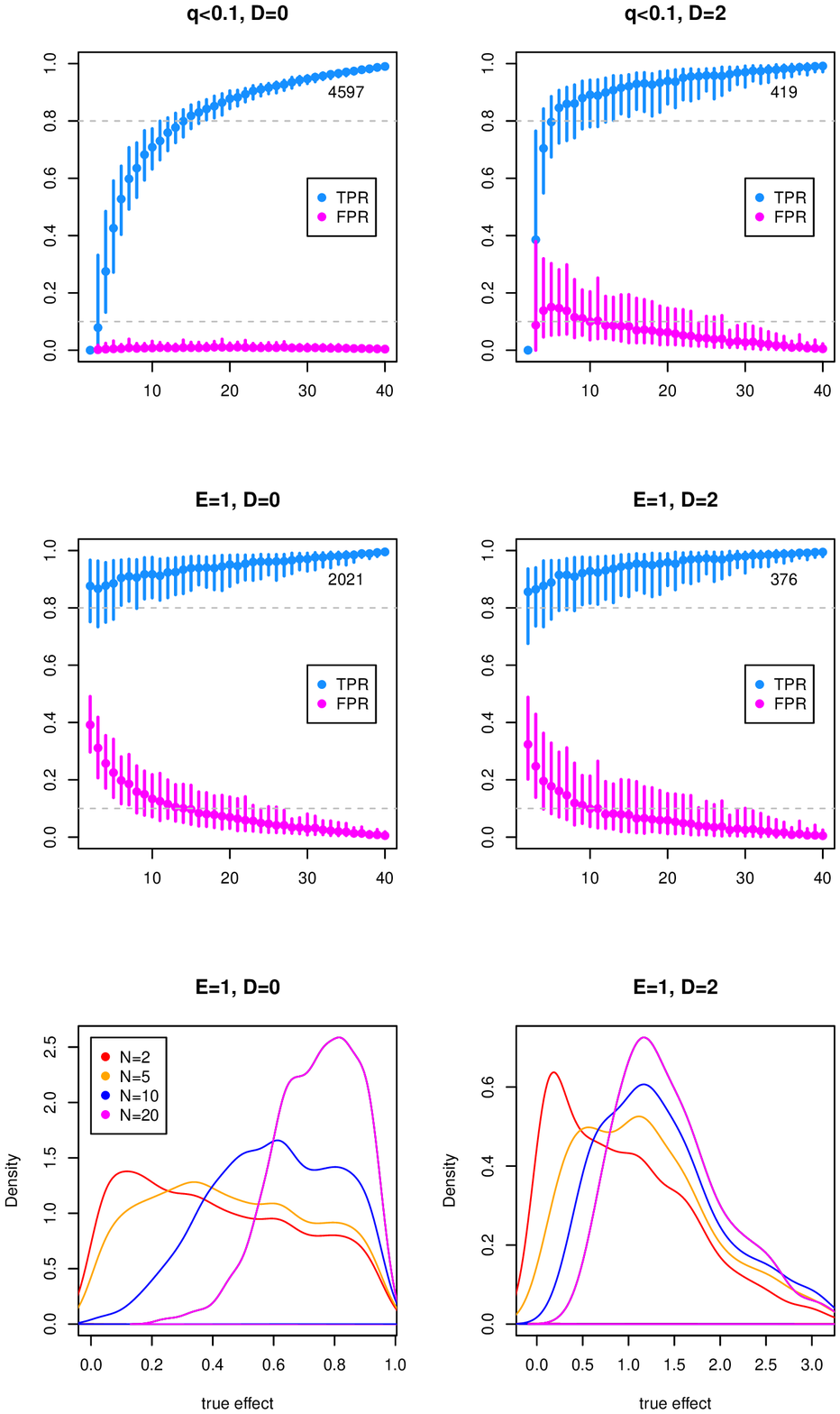}}
\caption{Comparing $\mathcal{E}_{d}$ and adjusted  p-values when detecting differential features between two groups. The top four panels show the   median and 95\% confidence interval of the true positive rate (TPR) and false positive rate (FPR) at different cutoff values.  The top two panels show the values for the q-value, the Benjamini-Hochberg corrected p-value, as a function of sample size. The middle two panels show the values for $\mathcal{E}_{d}$. Data were summarized using either a  0 or 2-fold difference cutoff (D). The number of features of the 6236 non-0 features that are identified as significantly different in the full dataset are shown in the top right corner of each plot. The bottom two panels show the effect size distribution of features identified as false positives by $\mathcal{E}_{d}$ at four different sample sizes. The dashed grey lines show the cutoff for a 10\% FDR and an 80\% power to detect.}
\label{fig:02}
\end{figure}

With this null behaviour information, we can examine an example dataset of a highly replicated RNA-seq dataset geneated by \citep{Schurch:2016aa}. In this dataset, the edgeR tool identified over 4600 out of 6349 genes as `significant'  (Benjamini-Hochberg adjusted p-value $< 0.05$) when all samples were included using either the glm or exact test modes (Supplementary Table 1).  Other widely used tools gave similar results \citep{Schurch:2016aa}. The null hypthesis testing framework in ALDEx2 also returned at least 4300 genes in the same dataset. Thus, identifying such a large proportion of genes as differentially abundant indicates that statistical significance is not informative for this type of experiment. Schurch et al. (and others) recommend adding a secondary threshold such as a fold-change cutoff to identify genes of interest for follow-up analyses \citep{Cui:2003aa,Schurch:2016aa}. When sample sizes are sufficiently large, we would expect that the fold-change cutoff itself would be the primary determinant of difference; however, this approach would not include either the biological variance or the uncertainty of measurement in the analysis.  

\begin{figure}[tpb]
\centerline{\includegraphics[scale=0.4]{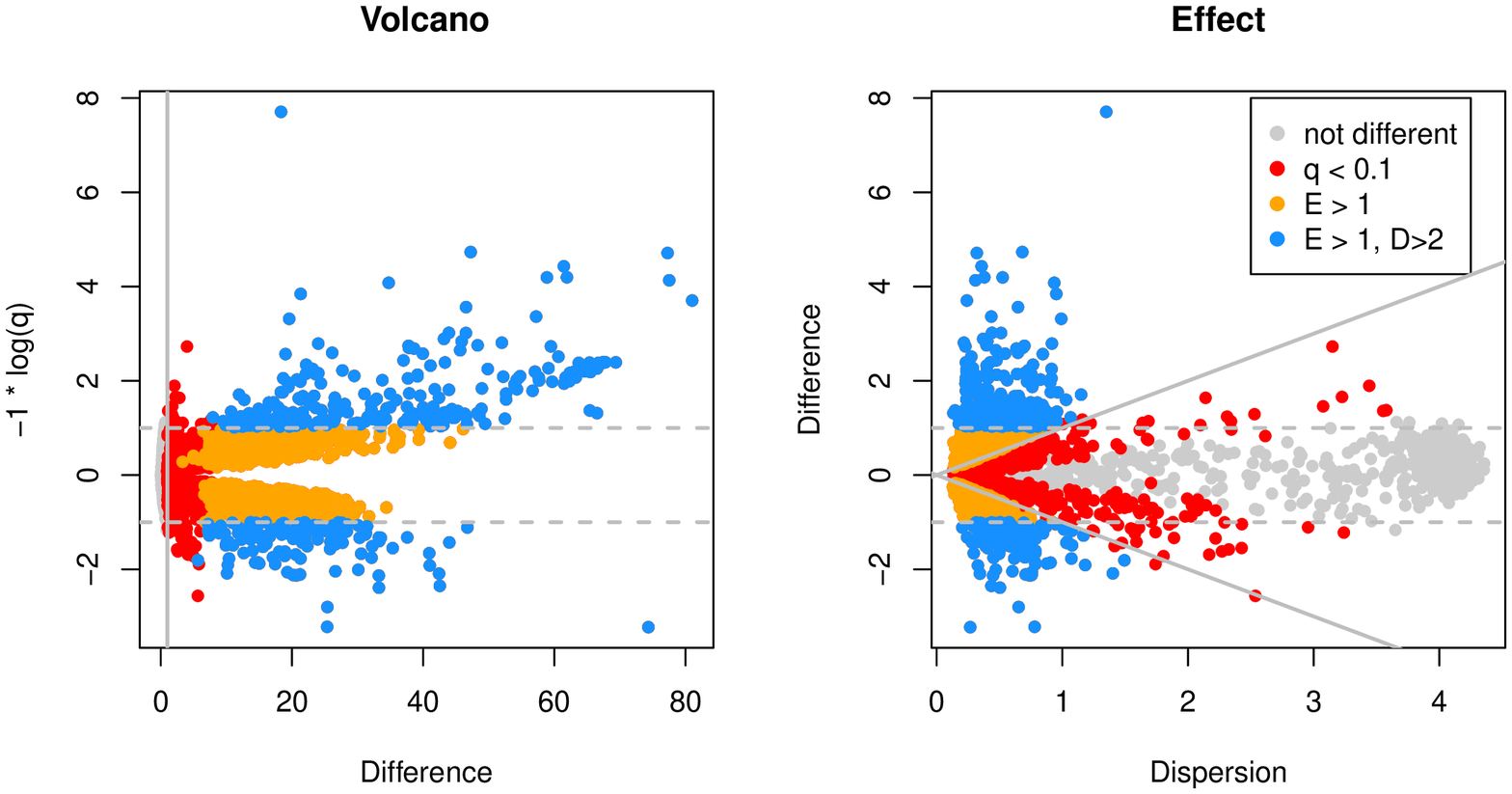}}
\caption{Volcano and Effect plots of features identified by  $\mathcal{E}_{d}$,  adjusted p-values  and absolute differences when detecting differential features between two groups. These plots compare the features identified by $\mathcal{E}_{d}$ and by q-scores (FDR) and a 2-fold fold-change thresholds in the full dataset.  In these plots all features that have a q score less than 0.1 also have an effect size greater than 1. Thus, the features in magenta are only identified as significantly different by q scores, those in orange are significantly different by both their q score and their effect size, and features in blue are significant by their q score, their effect size and their absolute difference.  The dashed grey lines in the two plots demarcate the 2-fold difference location; note that the difference is in a log2 scale. The Volcano plot is rotated from its usual orientation and the vertical solid line in the Volcano plot indicates a q score of 0.1. The diagonal solid lines in the Effect plot indicate the boundary where the difference equals the dispersion; ie, where the effect size is 1.
}
\label{fig:03}
\end{figure}

The relationship between sample size and the number of features identified as significantly different using a null hypothesis testing framework in this dataset is shown in the top  panels of Figure \ref{fig:02}.  Here we are testing for the ability to detect all features that would have been observed as differentially expressed in the full dataset when using a random subset of the data. The top two panels were generated using the expected value of the Benjamini-Hochberg corrected Welch's t-test by ALDEx2, but similar plots can be observed using the edgeR tool  or others \citep{Schurch:2016aa}. As expected we observe that the power of a t-test is strongly affected by sample size when the minimum absolute difference between groups is 0. However, when the minimum feature difference is a 2-fold change (D=2), the number of features identified is reduced approximately 10 fold, and a relatively small number of samples is required for acceptable power. The tradeoff here is that applying both the q-score and difference cutoffs results in an increase in the FDR at small numbers of samples. Note that all tools have difficulty estimating the actual FDR in many datasets \citep{Thorsen:2016aa,hawinkel2017}. 

The middle two panels of Figure \ref{fig:02} show the behaviour of the $\mathcal{E}_{d}$ statistic in the same random datasets. Note that even when only two samples are used, the $\mathcal{E}_{d}$ statistic identifies over 80\% of the features as different as are identified by the same statistic in the full dataset. Thus, the simple metric outlined here  can correctly identify the `true positive' set even when the number of samples is very small. The tradeoff when using this statistic is that at very low sample sizes the False Discovery Rate (fdr) is extreme; in this dataset and with and with a cutoff of $\mathcal{E}_{d} > 1$, the fdr is 40\% with two samples, but falls to less than 10\% only when there are 15 or more samples. Interestingly, applying a fold-change cutoff to the $\mathcal{E}_{d}$ metric reduces the false discovery rate dramatically and also reduces the number of features identified as significantly different. 

The bottom two panels of Figure \ref{fig:02} show the effect size in the full dataset of false positive features identified as different in subsets of the dataset. We can see that at a sample size of 2 the false positive features have true effect-sizes that are nearly randomly distributed, but that at a sample size of 10 or more, the vast majority of false positive features have true effect sizes that are at least 50\% of the desired effect size. When applying the difference cutoff, we observe that the majority of false positive features identified at small sample sizes have an effect size greater than the cutoff when the sample size is 5 or more. Thus, this implies that false positive features in this case pass the effect size cutoff but fail on the absolute difference cutoff. Supplementary Figure 6 shows a similar analysis for a 16S rRNA gene sequencing dataset \citep{bian:2017} with similar results, even thought he abundance/variance relationship between the features is very different than in the transcriptome experiment (Supplementary Figure 7). Supplementary Figure 8 shows that the effect thresholds from the simulated data in Figure 1 are appropriate, and perhaps even conservative, for real HTS data, and provide further evidence that the features identified as different in Figure 2 are likely to be reproducible.

Taking the data from Figures \ref{fig:01} and \ref{fig:02}, and the Supplement together, we can provide guidance as to appropriate cutoff values when using the $\mathcal{E}_{d} $ statistic in HTS datasets. First,  point estimates of $\mathcal{E}_{d} $ in real datasets and in simulated distributions are highly congruent. Thus, we can estimate that for every 100 features in a HTS dataset, we can use the curves in Figures \ref{fig:01} and  Supplementary Figure 8 to determine the number of false positive features expected if there is truly no difference between groups. The CoDaSeq R package contains a function that can be used to empirically determine thresholds for any desired percentile cutoff. As an example, the plots in Figure \ref{fig:01} that as a rule of thumb for the experimentalist to be 99\% confident in a true positive, effect sizes should be greater than 2 when the sample size is  4, greater than 1 when the sample size is 10, and greater than 0.5 for sample sizes larger than 40. This rule of thumb holds true regardless of the true underlying distribution and is appropriate whenever point estimates are computed. When computing the expected $\mathcal{E}_{d} $ as does the aldex.effect function, these thresholds are much lower at small sample sizes, and effect sizes can be about 30\% smaller. 

Figure \ref{fig:03} shows how the different threshold cutoffs relate to each other when plotted as a difference vs. q-score in a Volcano plot \citep{Cui:2003aa} and in a dispersion vs difference Effect plot \citep{gloor:effect}. These plots demonstrate the advantage of using a standardized effect metric over a q-score either with or without an absolute difference cutoff. In the Volcano plot, the q-score only features coloured in magenta, are exclusively features that have q scores near the upper bound of statistical significance. These include features with both large and small absolute differences, and the reason that a feature may have a large difference but a marginally significant q-score is not revealed on the Volcano plot. However, examination of the Effect plot shows that features that have a marginal q-score and large difference are features with very large dispersion; that is the difference between features, $\tilde{D}$, is much smaller than the within-group dispersion, MMAD, calculated as in Equation \ref{eq:ff}. Such features would not be expected to reproduce well in a new dataset because of their intrinsically high variance. In fact, these features are exactly those that are excluded by the $\mathcal{E}_{d} $ statistic.  Furthermore, adding the constraint that features have at least a 2-fold difference does not exclude these high-dispersion features from consideration. 

Examination of the orange features in the Volcano and Effect plots, we can see that both q-scores and $\mathcal{E}_{d} $ can exhibit arbitrarily small absolute differences if the dispersion is very small. In the case of the $\mathcal{E}_{d} $ statistic, adding in the requirement for at least a 2-fold change reduces the number of features to only those that are both different and reproducible.

By default, we want to know both `what is significant' and `what is different' \citep{Cui:2003aa}.  Both of these questions can be addressed with a standardized effect size statistic that scales the difference between features by their dispersion. We have found plots of difference and dispersion to be an exceeding useful tool when examining HTS datasets \citep{gloor:effect}. Furthermore, datasets analyzed by this approach have proven to be remarkably reproducible as shown by independent lab validation \citep{macklaim:2013, nelson:2015vaginal}

The $\mathcal{E}_{d}$ statistic outlined here is a relatively robust statistic with the attractive property that it consistently  identifies almost all the same set of true features regardless of the underlying distribution as shown in Figure \ref{fig:01}, and the number of samples as shown in Figure \ref{fig:02}. In marked contrast, even the best p-value based  approaches can identify only a small proportion of the features at small samples sizes that would have been found in the full dataset \citep{Schurch:2016aa}. Thus, the simple metric outlined here  can correctly identify the `true positive' set even when the number of samples is very small. Note that fold-change thresholds as is commonly used, is not the same as an standardized effect statistic, and applying the threshold values of \citep{Schurch:2016aa} while reducing the features that are found does not necessarily enhance reproducibility (Figures \ref{fig:02} and \ref{fig:03} ). 

The tradeoff when using the $\mathcal{E}_{d}$ statistic is that at very low sample sizes the False Discovery Rate can be extreme; in this dataset and with and with a cutoff of $\mathcal{E}_{d} > 1$, the FDR is 40\% with two samples, but falls to less than 10\% only when there are 15 or more samples. Adding in an absolute fold-change restriction reduces the FDR substantially. Further tempering this, is the observation that the false positive features are frequently  close to the cutoff in the full dataset: that is, false positive features typically are true positives at slightly lower effect sizes or absolute fold changes. This is in contrast to the well-known random uniform distribution of false positive p-values. The Supplement shows additional evidence that the  $\mathcal{E}_{d}$ statistic is generally useful, having essentially the same characteristics in a 16S rRNA gene sequencing dataset which has much larger per feature dispersion. 

This work  describes the  $\mathcal{E}_{d}$ statistic for examining high throughput sequencing datasets. The statistic is relatively robust and efficient, and answers the question most desired by the biologist, namely `what is reproducibly different'.   $\mathcal{E}_{d}$ is computed in the ALDEx2 R package as the `effect' output where it is the median of the inferred technical and biological data, and in the CoDaSeq R package where it acts only the point estimates of the data. Interactive exploration of effect sizes can be done in the omicplotR Bioconductor package \citep{omicplot}.

\section*{Authors contributions}
AF devised the $\mathrm{d_{NEF}}$ metric. MTHQV and LM did simulation studies to characterize the metric. JM characterized the properties of the metric in real data. GBG characterized the metric, did final simulation studies, wrote the first draft manuscript and supplement. All authors provided input into the final form and substance of the manuscript.

\section*{References}
\bibliographystyle{unsrtnat}

\end{document}